\documentclass[submission, Phys]{SciPost}

\hypersetup{
    colorlinks,
    linkcolor={red!50!black},
    citecolor={blue!50!black},
    urlcolor={blue!80!black}
}
\usepackage{comment}
\usepackage{upgreek}
\usepackage{academicons}
\usepackage{xcolor}
\usepackage{svg}
\usepackage{orcidlink}
\usepackage[bitstream-charter]{mathdesign}
\DeclareSymbolFont{usualmathcal}{OMS}{cmsy}{m}{n}
\DeclareSymbolFontAlphabet{\mathcal}{usualmathcal}

\begin{document}

\begin{center}{\Large \textbf{
Status and Prospects of the JUNO Experiment\\
}}\end{center}

% TODO: write the author list here. Use initials + surname format.
% Separate subsequent authors by a comma, omit comma at the end of the list.
% Mark the corresponding author with a superscript *.
\begin{center}
Matthias Raphael Stock\textsuperscript{1$\star$}\orcidlink{0000-0002-5963-7431} on behalf of the JUNO collaboration
\end{center}

\begin{center}
{\bf 1} Technical University of Munich, TUM School of Natural Sciences, Department of Physics, James-Franck-Str. 1, 85748 Garching, Germany
\\
* raphael.stock@tum.de
\end{center}

\begin{center}
\today
\end{center}

% For convenience during refereeing (optional),
% you can turn on line numbers by uncommenting the next line:
%\linenumbers

\definecolor{palegray}{gray}{0.95}
\begin{center}
\colorbox{palegray}{
  \begin{tabular}{rr}
  \begin{minipage}{0.1\textwidth}
    \includegraphics[width=30mm]{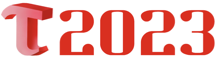}
  \end{minipage}
  &
  \begin{minipage}{0.81\textwidth}
    \begin{center}
    {\it The 17th International Workshop on Tau Lepton Physics}\\
    {\it Louisville, USA, 4-8 December 2023} \\
    \doi{10.21468/SciPostPhysProc.?}\\
    \end{center}
  \end{minipage}
\end{tabular}
}
\end{center}

\section*{Abstract}
{\bf
% TODO: write your abstract here.
%The abstract is in boldface, and should fit in 8 lines.
%It should be written in a clear and accessible style, emphasizing the context, the problem(s) studied, the methods used, the results obtained, the conclusions reached, and the outlook. %You can add a table contents, recommended if your paper is more than 6 pages long.
%\\ \\
The Jiangmen Underground Neutrino Observatory~(JUNO) is a multi-purpose neutrino experiment currently under construction in China. It is located~52.5\,km away from two nuclear power plants in a newly constructed~700-m-deep underground laboratory.
JUNO will be the largest liquid scintillator~(LS) detector in the world comprising 20\,kt of ultra-pure LS filled in an acrylic sphere.
Its main goal is to determine the neutrino mass ordering by measuring the energy spectrum of reactor neutrinos with highest accuracy.
In addition,~JUNO will cover precision measurements of oscillation parameters and several aspects in the field of astroparticle physics.
Data taking will start in late~2024.
}

%\vspace{10pt}
%\noindent\rule{\textwidth}{1pt}
%\tableofcontents\thispagestyle{fancy}
%\noindent\rule{\textwidth}{1pt}
%\vspace{10pt}

\section{Introduction}
\label{sec:intro}
% TODO: write your article here.
%The stage is yours. Write your article here.
%The bulk of the paper should be clearly divided into sections with short descriptive titles, including an introduction and a conclusion.

\begin{figure}[h]
	\centering
	\includegraphics[width=0.475\textwidth]{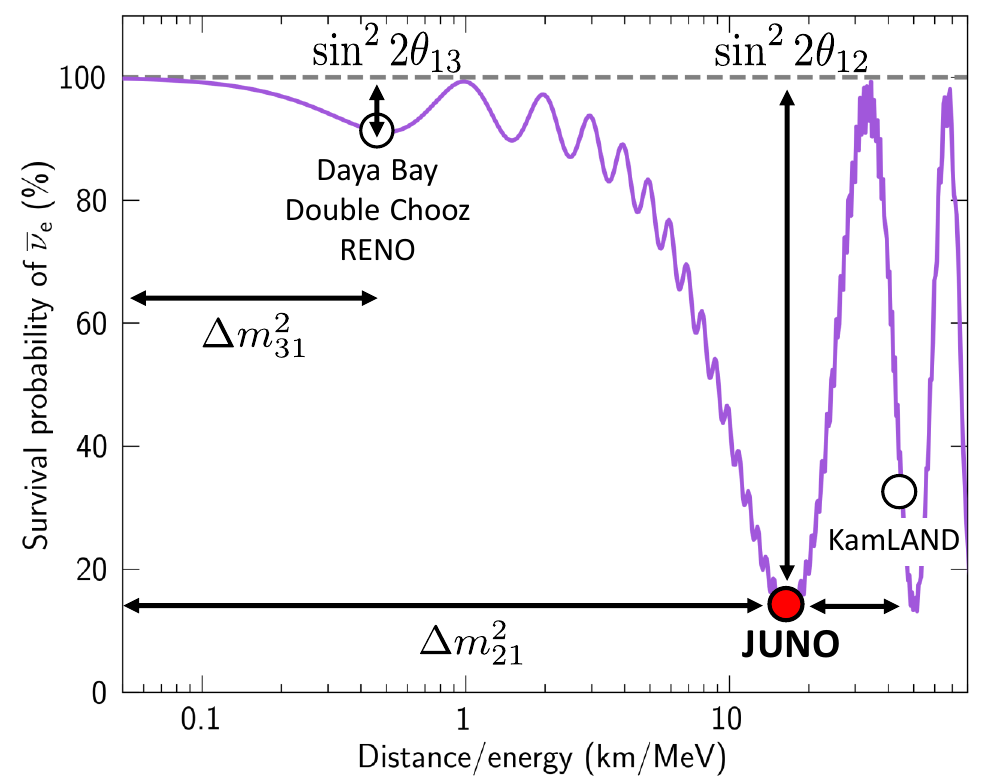}
	\includegraphics[width=0.475\textwidth]{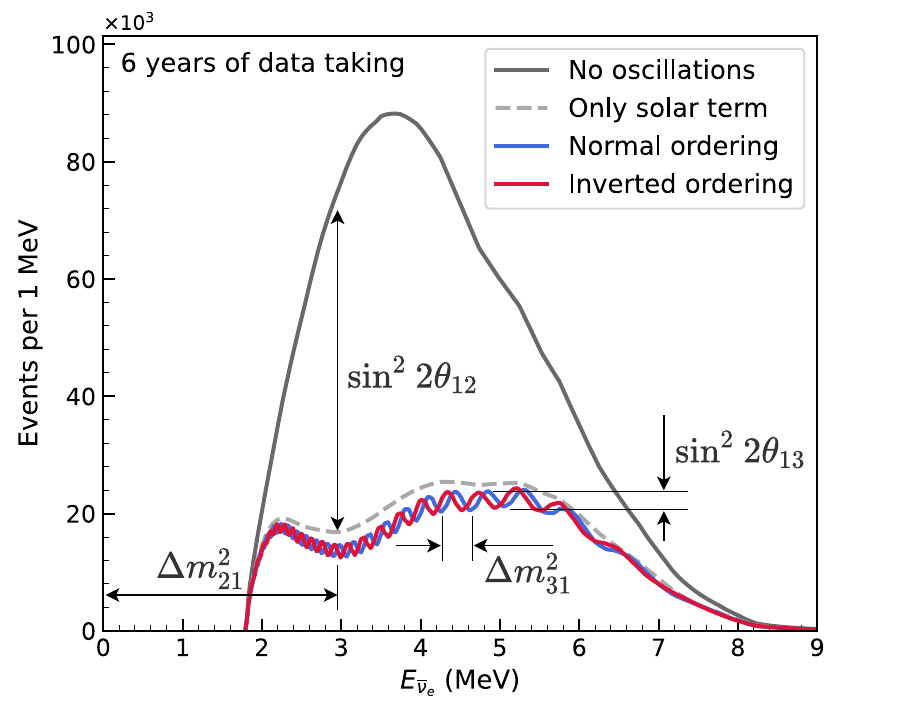}
	\caption{\textbf{Left:} JUNO is located where the survival probability of the reactor neutrinos is minimal. \textbf{Right:} Expected reactor neutrino energy spectra are illustrated assuming perfect detector resolution of~JUNO and six years of data taking~\cite{JUNOSubPercentPrecisionOsciParam2022}.}\label{fig:SurvEnergySpec}
\end{figure}

One of the most important current missions in neutrino physics is to determine the neutrino mass ordering~(NMO). Since neutrino oscillations are enhanced by the matter effect, the positive mass splitting of~$\Delta m^2_{21} = m^2_2 - m^2_1 > 0$ is observed by solar neutrino experiments. The question remains whether the third mass eigenstate~$m_3$ is the heaviest or the lightest leading to two possible scenarios of normal ordering~(NO) and inverted ordering~(IO), respectively. 
The primary goal of the Jiangmen Underground Neutrino Observatory~(JUNO) is to determine the~NMO by studying the oscillation pattern of reactor neutrinos. JUNO is currently under construction in the Guangdong province in South China. 
It is located in a newly constructed 700-m-deep underground laboratory~(1,800\,m.w.e.) at an average distance of~52.5\,km to eight reactor cores of the two nearest nuclear power plants Yangjiang and Taishan, which have a combined thermal power of~26.6\,GW$_{\text{th}}$~\cite{JUNOSubPercentPrecisionOsciParam2022}.
The medium-baseline is chosen because the survival probability of electron antineutrinos~$\overline{\upnu}_{\text{e}}$ emitted from the reactor cores is minimal as Figure~\ref{fig:SurvEnergySpec}~(Left) illustrates. Here, the ‘slow solar oscillation’ pattern causes the global minimum and depends on~$\Delta m^2_{21}$ and~$\sin^2 2\theta_{12}$ while the ‘fast atmospheric oscillation’ pattern causes the superimposed ripples and depends on~$\Delta m^2_{31}$ and~$\sin^2 2\theta_{13}$. Therefore, this optimal location enables the first-time precision measurement of two oscillation patterns in one energy spectrum.
Compaired to long-baseline experiments, the advantage of reactor experiments is that the survival probability is vacuum-dominated and independent of~the mixing angle~$\theta_{23}$ and~the~CP-violating phase~$\delta_{\text{CP}}$.
Figure~\ref{fig:SurvEnergySpec}~(Right) illustrates the expected energy spectra after six years of data taking. Absolute value and sign of~$\Delta m^2_{31}$ lead to the different position of the ripples in the energy spectra of~NO and~IO.
JUNO can distinguish between the two scenarios and reveal the realized~NMO in Nature by recording large statistics and in high precision. 

\section{Detector Design and Status}

\begin{figure}[h]
	\centering
	\includegraphics[width=0.7335\textwidth]{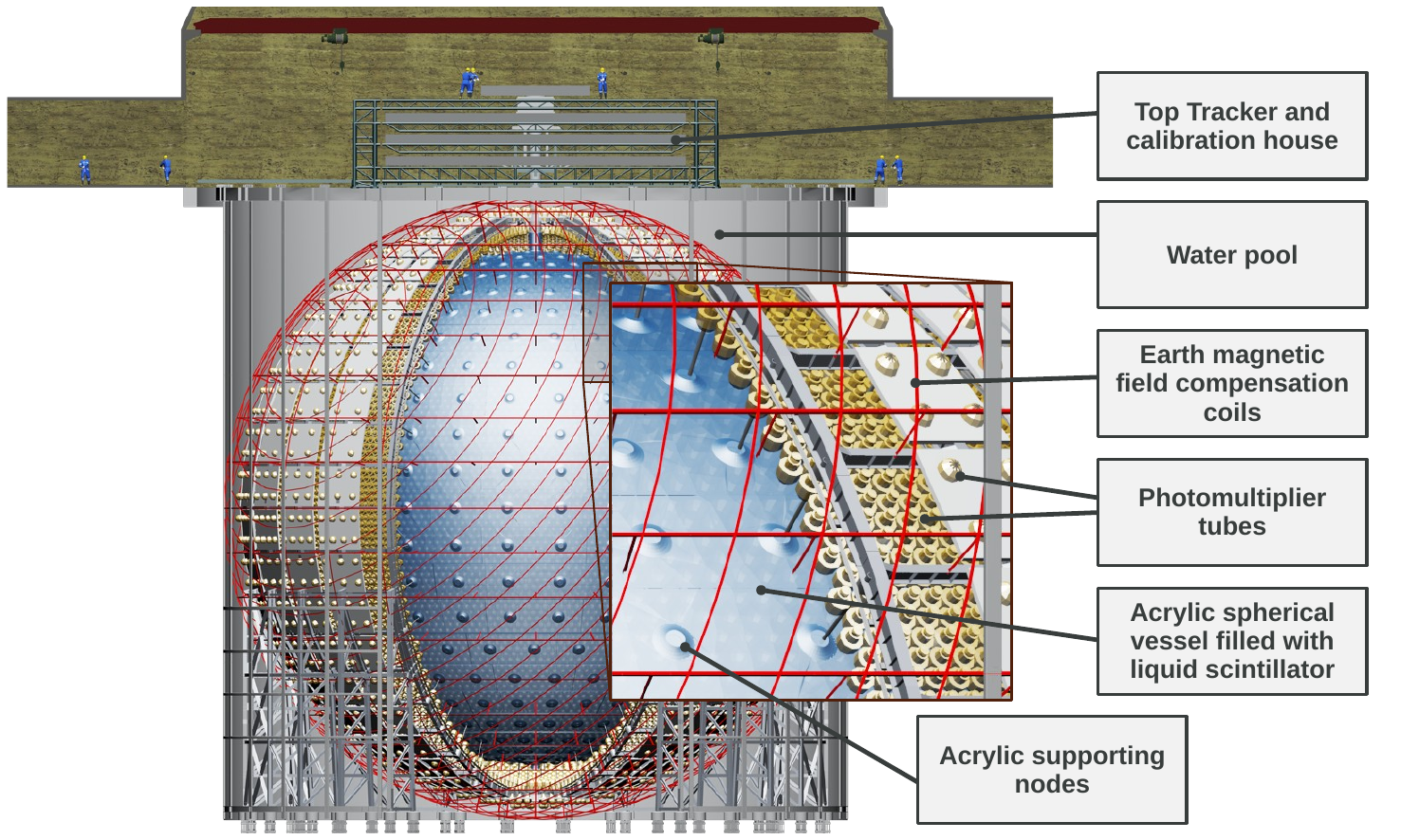}
	\caption{Overview of the~JUNO detector~\cite{JUNOSubPercentPrecisionOsciParam2022}.}
	\label{fig:JUNO}
\end{figure}

The ability to resolve the~NMO sets demanding design requirements for the detector. Civil construction of the JUNO site was completed in December~2021.
JUNO will be the largest liquid scintillator~(LS) detector in the world comprising~20\,kt of ultra-pure LS filled in a~35.4-m-diameter acrylic sphere.  
The vast volume is monitored by~17,612 large~20-inch and~25,600 small~3-inch photomultiplier tubes, denoted as~LPMT and~SPMT, respectively. They are mounted on a stainless steel support structure that was finished in June~2022 followed by the assembly of acrylic panels forming the sphere. Figure~\ref{fig:JUNO} presents an overview of the~JUNO detector. The central detector~(CD) ensures a photocathode coverage of~78\,\% and provides an unprecedented energy resolution of better than~3\,\% at~1\,MeV. The~CD is surrounded by the~43.5-m-diameter cylindrical water Cherenkov detector~(WCD) comprising~35\,kt of ultra-pure water viewed by~2,400~LPMTs. The~WCD provides a veto for cosmic muons and shields the~CD from environmental radioactivity~\cite{YellowBookJUNO, JUNOphysicsanddetector2022, JUNOSubPercentPrecisionOsciParam2022}.
The Top Tracker~(TT) covers~$60\,\%$ of the~CD and~WCD and measures~30\,\% of cosmic muons passing through the~CD with a median resolution of~0.2\,\% at the bottom of the~WCD. Combined with the~CD and~WCD, the~TT provides well reconstructed muon samples with~$>99\,\%$ purity to calibrate and tune reconstruction algorithms and veto affected regions~\cite{JUNOphysicsanddetector2022, JUNO_TopTracker2023}.
Several subsystems use radioactive sources or pulsed~UV light to regulary calibrate the energy scale, which is necessary due to the non-linearity of the~LS~\cite{JUNOCalibrationStrategy2021}. The precision of the absolute energy scale better than~1\,\% is guaranteed~\cite{YellowBookJUNO, JUNOphysicsanddetector2022, JUNOSubPercentPrecisionOsciParam2022}.

\subsection{Liquid Scintillator Purification and OSIRIS}

\begin{figure}[h]
	\centering
	\includegraphics[width=0.835\textwidth]{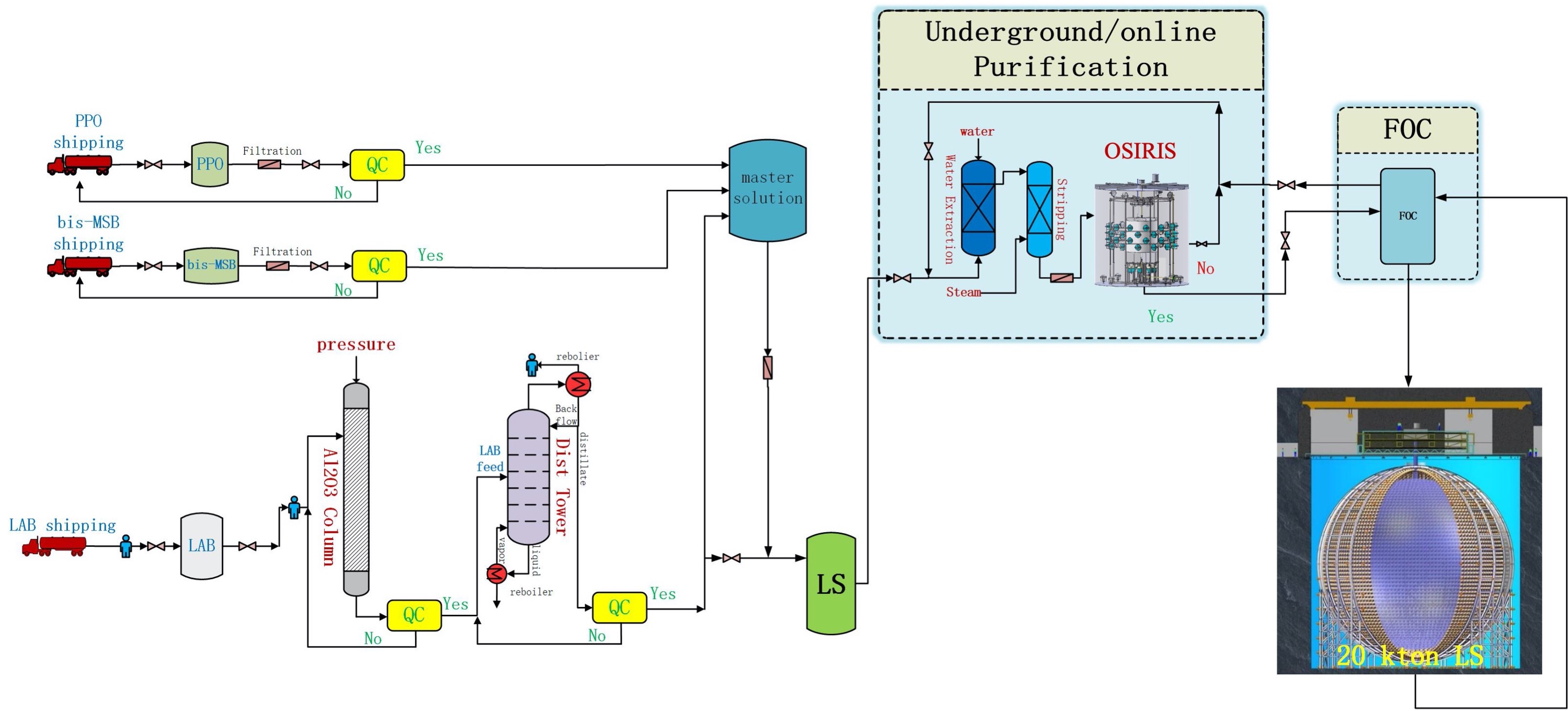}
	\caption{Overview of the liquid scintillator systems with purification chains above ground and underground including~the pre-detector OSIRIS%. Figure taken from
	~\cite{JUNOphysicsanddetector2022}.}
	\label{fig:LSsystem}
\end{figure}

The liquid scintillator~(LS) of~JUNO is composed of~LAB~(linear alkylbenzene) as the solvent mixed with~2.5\,g/L~PPO~(2,5-diphenyloxazole) as fluor and~3\,mg/L~bis-MSB~(1,4-bis(2-methylstyryl)benzene) as additional wavelength shifter~\cite{JUNO_LS}. Optical properties are crucial and require,~e.g., very high light yield of about~10,000 photons per MeV and an attenuation length~$\gtrsim 20\,$m at~430\,nm due to the size of the~CD. Radiopurity is crucial requiring~U/Th concentrations of~$\lesssim 10^{-15}$\,g/g for the~NMO determination and~$\lesssim 10^{-17}$\,g/g for solar neutrino measurements. Therefore, the~LS passes a chain of purification steps onsite as~Figure~\ref{fig:LSsystem} shows. Before the~LS mixing, the~LAB goes through the~Al$_{2}$O$_3$ column to remove optical impurities. The distillation plant removes heavy and high-boiling radioactive materials such as~$^{238}$U,~$^{232}$Th and~$^{40}$K. Then, the~LS will be mixed with~PPO and~bis-MSB, pumped underground where water extraction and steam stripping are applied to purify further and remove gaseous impurities, respectively~\cite{Plants}. The quality of a fraction of the produced~LS is checked by the Online Scintillator Internal Radioactivity Investigation System~(OSIRIS) before and during the filling of the~CD.
OSIRIS acts as a pre-detector and verifies the radiopurity and optical properties of the~LS~\cite{OSIRIS2021}. The purification plants and~OSIRIS are currently under commissioning.


\subsection{Taishan Antineutrino Observatory}

The Taishan Antineutrino Observatory~(TAO)~\cite{TAOCDR2020} is a satellite detector of~JUNO located~$44$\,m away from the Taishan reactor core~1. Its purpose is to measure the unoscillated antineutrino energy spectrum with extremely high statistics and in unprecedented high-resolution, which is smaller than~2\,\% at~1\,MeV. This model-independent reference spectrum reduces the shape uncertainty and improves the sensitivity of JUNO for the~NMO. Besides,~TAO provides benchmark measurements to test nuclear databases. It also allows to search for short neutrino oscillations caused by hypothetical coupling of sterile neutrinos to the flavor eigenstates violating the unitarity of the usual~$3\times3$ mixing matrix.~TAO uses novel detector technologies,~e.g.,~10\,m$^2$ of~silicon photomultipliers~with~$50\,\%$ detection efficiency achieve an almost full optical coverage of the~2.8\,t~gadolinium-loaded~LS volume operated at~$-50^\circ$C~\cite{TAOCDR2020}. The~TAO detector has been assembled and tested at the~IHEP and will be moved to the Taishan site in~2024.

\section{Physics Prospects}

JUNO will have a broad and rich physics program due to its large detector target mass, excellent energy resolution, outstanding radiopurity and good shielding. Beyond the determination of the NMO,~JUNO will measure several neutrino mixing parameters with sub-percent precision, significantly contribute to the field of astroparticle physics and search for physics beyond the Standard Model of Particle Physics~\cite{YellowBookJUNO, JUNOphysicsanddetector2022}.
A few examples are highlighted in the following.

\subsection{Neutrino Mass Ordering and Neutrino Oscillation Parameters}

\begin{figure}[h]
	\centering
	\includegraphics[width=0.495\textwidth]{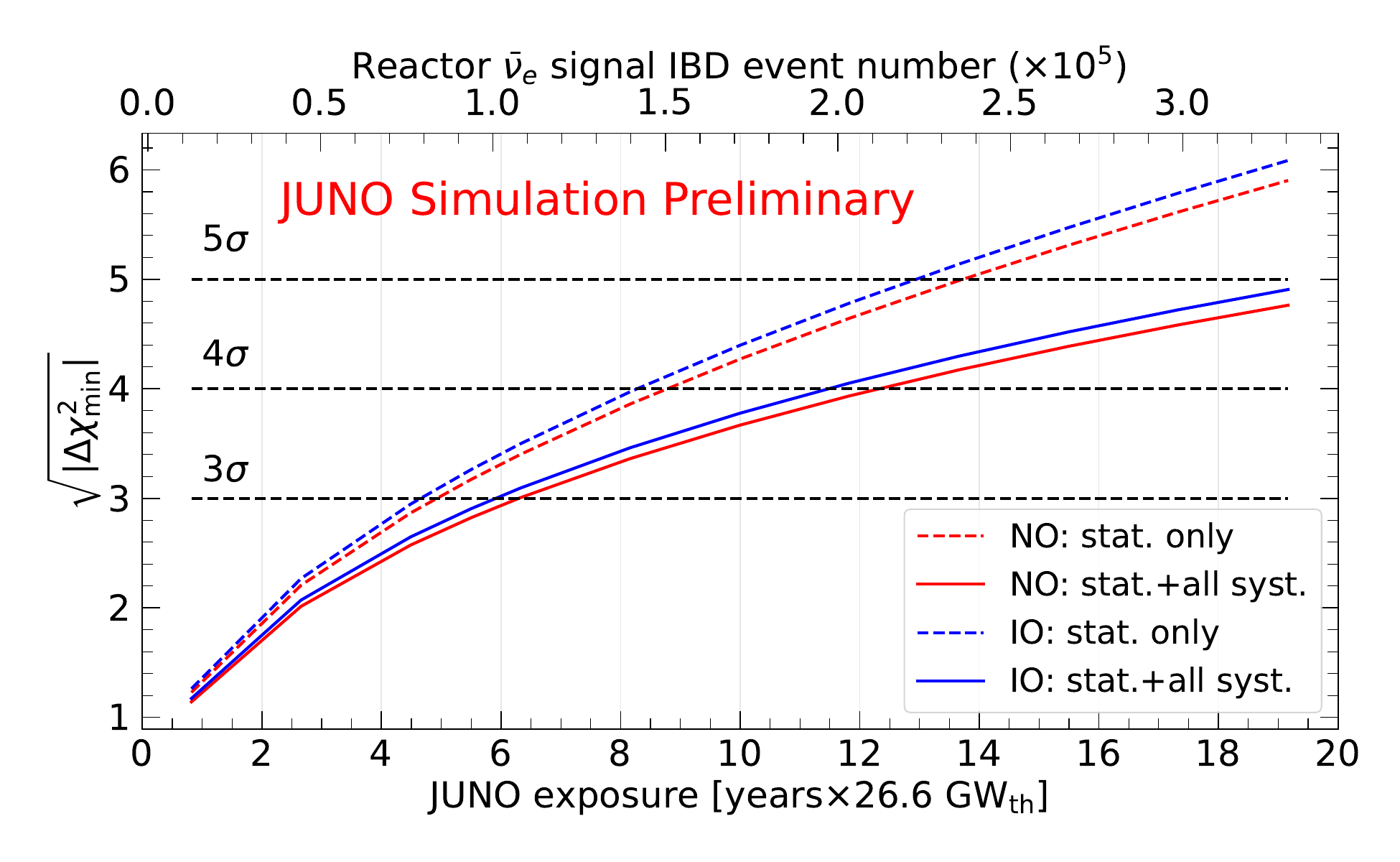}	\includegraphics[width=0.495\textwidth]{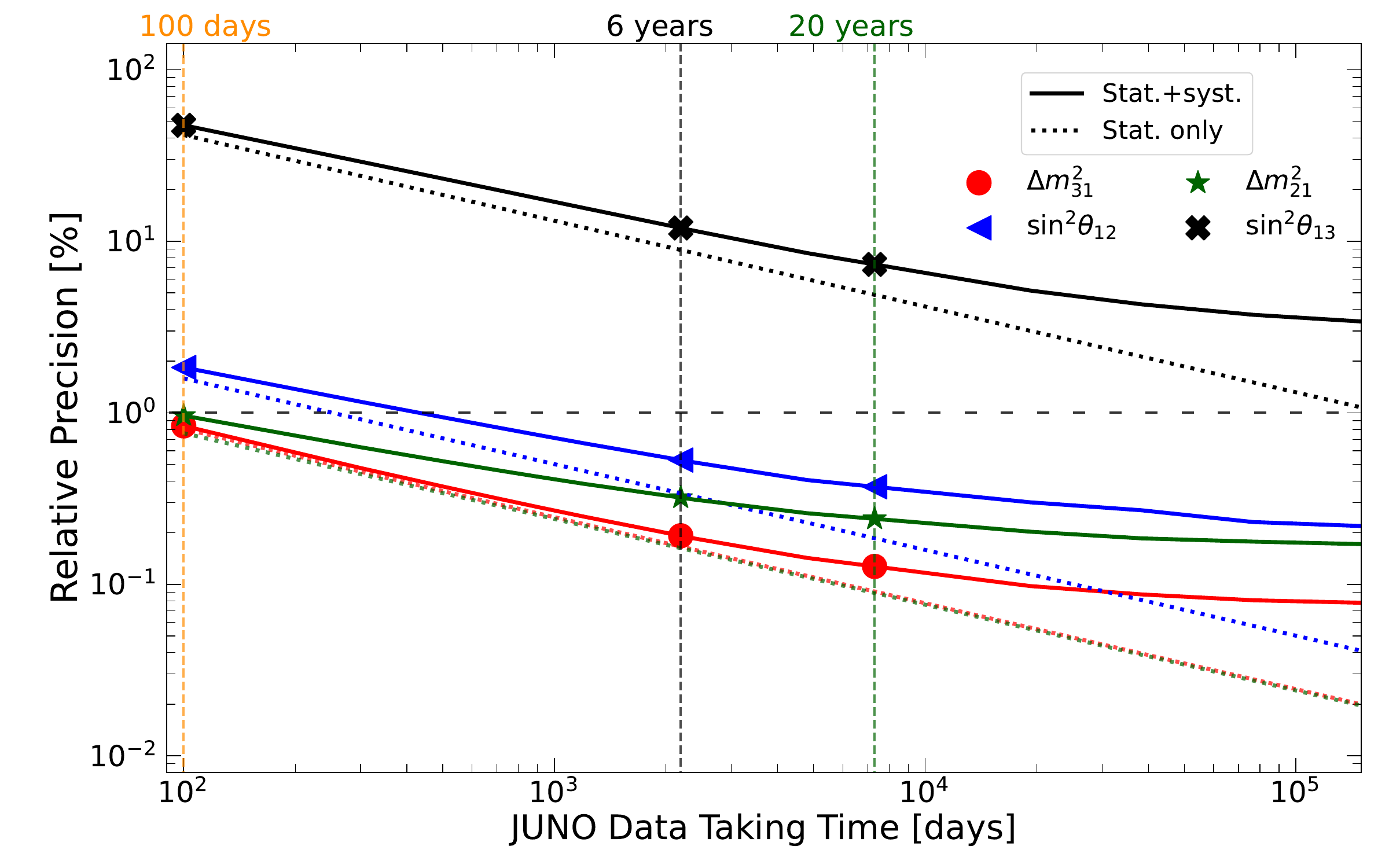}
	\caption{\textbf{Left:} Evolution of sensitivity for~NO and~IO~\cite{ZhaoNeutrino2022}. \textbf{Right:} Evolution of relative precision for the oscillation parameters~$\Delta m^2_{31}$,~$\Delta m^2_{21}$,~$\sin^2\theta_{12}$ and~$\sin^2\theta_{13}$~\cite{JUNOSubPercentPrecisionOsciParam2022}.}
	\label{fig:NMOandOsciParams}
\end{figure}

JUNO measures~approximately~$45$ reactor neutrino events per day via the inverse beta dacay~(IBD), $\overline{\upnu}_{\text{e}} + \text{p} \rightarrow \text{e}^{+} + \text{n}$, which has an energy threshold of~$E_\text{thr} = 1.8\,\text{MeV}$ and features a time-correlated scintillation signature in the detector.
The prompt signal is caused by the scintillation of the positron~e$^+$ and its annihilation producing two gammas of~$E_\upgamma =0.511\,\text{MeV}$ each exciting the~LS. The antineutrino energy~$E_{\overline{\upnu}_{\text{e}}}$ can be reconstructed from the energy deposition of the positron,~i.e.,~$E_{\text{prompt}} = 2\times E_\upgamma + E_{\overline{\upnu}_{\text{e}}} - E_\text{thr} = %\simeq
E_{\overline{\upnu}_{\text{e}}} - 0.8\,\text{MeV}$. After the thermalization of the neutron, it is captured by a proton releasing~2.2\,MeV in gamma radiation, which exites the~LS and causes the delayed signal.
The measured energy spectrum of the reactor neutrinos will be fit assuming~normal ordering~(NO) or inverted ordering~(IO). The difference of the minimal chi-square values between both fits is defined as the discriminator of the~NMO,~i.e.,~$\Delta \chi^2_{\text{min}}$.
Figure~\ref{fig:NMOandOsciParams}~(Left) shows the expected sensitivity evolution to the~NMO for both scenarios of~NO and~IO including statistical and systematic uncertainties. Therefore,~JUNO can determine the~NMO at a~$3\,\sigma$ confidence level after about six years of data taking~\cite{JUNOphysicsanddetector2022, ZhaoNeutrino2022}. 

Besides, the sensitivity of~JUNO for measuring several neutrino oscillation parameters has been investigated in~\cite{JUNOSubPercentPrecisionOsciParam2022}.
It is expected that JUNO will measure the two ‘solar’ parameters,~i.e.,~$\Delta m^2_{21}$ and~$\sin^2\theta_{12}$, and the ‘atmospheric’ mass splitting~$\Delta m^2_{31}$ with sub-perccent precision after six years of data taking as Figure~\ref{fig:NMOandOsciParams}~(Right) shows. It is also predicted that~JUNO will already cross the current world-leading precision after the first~100\,days~\cite{JUNOSubPercentPrecisionOsciParam2022}. Therefore,~JUNO will contribute significantly to global tests of the unitarity of the neutrino mixing matrix.

\subsection{Geoneutrinos}

	Geoneutrinos originate from radioactive decay chains of naturally abundant and long-lived isotopes inside Earth such as~$^{238}$U and~$^{232}$Th. The observation of geoneutrinos is a unique and non-invasive tool to study,~e.g., the amount of radiogenic heat, the chemical composition or formation processes of the Earth.
	JUNO will measure around~400 events per year via~IBD significantly improving the statistics of the existing global geoneutrino event sample~\cite{YellowBookJUNO, JUNOphysicsanddetector2022}.

\subsection{Solar Neutrinos}

	Solar neutrinos provide valuable information,~e.g., about the solar metallicity or neutrino oscillations. They are produced as electron neutrinos along reaction chains of nuclear fusion processes in the core of the Sun. 
	JUNO measures solar neutrinos via elastic neutrino scattering~(ES),~$\upnu_{\text{e}} + \text{e}^- \rightarrow \upnu_{\text{e}} + \text{e}^-$ and they are studied via neutrino spectroscopy.
	Sensitivity studies of~$^8$B solar neutrinos predict a precision of~5\,\%,~8\,\% and~20\,\% for their flux,~$\sin^2\theta_{12}$ and~$\Delta m^2_{21}$, respectively~\cite{JUNO_Solar8B_2021, JUNO_SolarModelIndependent8B_2022}.
	Sensitivity studies of~$^7$Be,~pep and~CNO solar neutrinos were conducted assuming several radiopurity scenarios of the~LS. The precision on the fluxes can be improved in most cases~\cite{JUNO_SolarNeutrinos_2023}.

\subsection{Supernova Neutrinos and Multi-Messenger Trigger System}
	
	The observation of supernova burst neutrinos is a unique opportunity to study,~e.g., stages and mechanisms of supernova~(SN) explosions or search for collective neutrino oscillations.
	Assuming one nearby core-collapse supernova~(CCSN) at a distance of~10\,kpc,~JUNO will detect large statistics of SN burst neutrinos in multiple-flavor detection channels,~i.e., about~5,000~IBD events, about~2,000 all-flavor neutrino-proton ES events via~$\upnu + \text{p} \rightarrow \upnu + \text{p}$, and about~300 neutral-current~(NC) ES signals while~CC and~NC interactions on~$^{12}$C nuclei are also observable~\cite{JUNOphysicsanddetector2022, JUNO_Supernova_2023}.
	Besides, an independent multi-messenger~(MM) trigger system provides an ultra-low detection threshold of few tens of keV allowing to observe low-energy transient signals and to monitor~CCSNe with an extended energy band.
	JUNO will play a major role in the global network of~MM observatories and in the next-generation Supernova Early Warning System~\cite{JUNOphysicsanddetector2022, JUNO_Supernova_2023}.
	Another objective is the search of the diffuse supernova neutrino background~(DSNB), which is the cumulative neutrino flux from past supernovae in the visible Universe.
	The observation would provide information about the average~CCSN energy spectrum, the cosmic star-formation rate or the fraction of black hole formation. 
	JUNO could identify a few~DSNB events per year via~IBD within an energy window between around~$10$\,MeV to~$30$\,MeV defined by the~IBDs of reactor and atmospheric neutrinos. Main background within this window are the~NC interactions of atmospheric neutrinos, which can be effectively reduced by applying pulse shape discrimination. 
	The discovery potential of the~DSNB is estimated to~$3\,\sigma$ after three years of data taking while non-observation would significantly improve current limits and constrain the model parameter space~\cite{JUNOphysicsanddetector2022, JUNO_DSNB2022}.

\subsection{Atmospheric Neutrinos}
	JUNO measures several atmospheric neutrinos per day below and within the~GeV  range. The reconstruction of atmospheric electron and muon neutrino energy spectra is expected to reach~10\,\%~to~25\,\% precision with five years of data.
	Beyond that, detecting oscillated atmospheric neutrinos	allows to investigate the octant of the mixing angle~$\theta_{23}$ and to enhance the sensitivity of~JUNO for the~NMO by providing complementary inputs~\cite{JUNOphysicsanddetector2022, JUNOSensitivityLowAtmospheric2021}. Besides,~JUNO will be able to detect the appearance of~tau neutrinos~$\upnu_\uptau$~\cite{SMB}.

\subsection{Proton Decay}
	
	The observed matter-antimatter asymmetry in the Universe can be explained by baryon number violation, which is an inevitable consequence of Grand Unified Theories.
	JUNO is able to probe the possible proton decay channel~$\text{p} \rightarrow \overline{\upnu} + \text{K}^+$, which features a clear three-fold time-correlated scintillation signature due to the prompt signal of the kaon, a short delayed signal of its daughter and the late signal of the daughter's decay positron.
	JUNO will be competitive to world-leading lower limits on the proton lifetime by being sensitive to~$>9.6 \times 10^{33}\,\text{years}$ at~90\,\%~C.L. after~10 years of data taking~\cite{JUNOphysicsanddetector2022, JUNO_ProtonDecay_2023}.

\section{Conclusion}

JUNO is a next-generation large-scale neutrino observatory whose main goal is to determine the neutrino mass ordering with a~$3\,\sigma$ significance after six years of data taking. Among other things, the study of several neutrino properties and neutrino sources, including the Sun, the Earth’s interior, the atmosphere or core-collapse supernovae complement the rich physics program. 
The filling of the~JUNO detector and the start of data taking are expected before the end of~2024.

\section*{Acknowledgements}
M.R.S. wishes to thank the organizers of this conference and the JUNO collaboration for the opportunity and invitation to give this presentation.
The JUNO collaboration is grateful for the ongoing cooperation from the China General Nuclear Power Group.
% TODO: include author contributions
%\paragraph{Author contributions}
%This is optional. If desired, contributions should be succinctly described in a single short paragraph, using author initials.

% TODO: include funding information
\paragraph{Funding information}
M.R.S. is supported by the German Collaborative Research Center “Neutrinos and Dark Matter in
Astro- and Particle Physics”~(SFB 1258) and the Deutsche Forschungsgemeinschaft~(DFG) (FOR 5519).
This work was supported by the Chinese Academy of Sciences, the National Key~R\&D Program of China, the CAS Center for Excellence in Particle Physics, Wuyi University, and the Tsung-Dao Lee Institute of Shanghai Jiao Tong University in China, the Institut National de Physique Nucléaire et de Physique de Particules~(IN2P3) in France, the Istituto Nazionale di Fisica Nucleare~(INFN) in Italy, the Italian-Chinese collaborative research program MAECINSFC, the Fond de la Recherche Scientifique (F.R.S-FNRS) and~FWO under the “Excellence of Science — EOS” in Belgium, the Conselho Nacional de Desenvolvimento Científico e Tecnològico in Brazil, the Agencia Nacional de Investigacion y Desarrollo and~ANID — Millennium Science Initiative Program — ICN2019\_044 in Chile, the Charles University Research Centre and the Ministry of Education, Youth, and Sports in Czech Republic, the Deutsche Forschungsgemeinschaft~(DFG), the Helmholtz Association, and the Cluster of Excellence PRISMA$^+$ in Germany, the Joint Institute of Nuclear Research~(JINR) and Lomonosov Moscow State University in Russia, the joint Russian Science Foundation~(RSF) and National Natural Science Foundation of China (NSFC) research program, the MOST and MOE in Taiwan, the Chulalongkorn University and Suranaree University of Technology in Thailand, University of California at Irvine and the National Science Foundation in~U.S.A.
\bibliography{SciPost_Example_BiBTeX_File.bib}

\nolinenumbers

\end{document}